\documentclass{revtex4}

\usepackage{graphicx}
\def\eslt{\not\!\!{E_T}}
\def\to{\rightarrow}

\def\tb{\tilde b}

\def\tst{\tilde t}

\def\tg{\tilde g}

\def\tw{\widetilde\chi^\pm}

\def\tz{\widetilde\chi^0}

\begin{document}
\bibliographystyle{revtex}


\title{The complementary roles of the LHC and the LC 
in discovering supersymmetry}

\author{Howard Baer}
\email[]{baer@hep.fsu.edu}
\author{Csaba Bal\'azs}
\email[]{balazs@hep.fsu.edu}
\affiliation{Department of Physics,
Florida State University,
Tallahassee, FL 32306 USA}
\author{J.~Kenichi Mizukoshi}
\email[]{mizuka@phys.hawaii.edu}
\affiliation{Department of Physics and Astronomy,
University of Hawaii, Honolulu, HI 96822, USA}


\begin{abstract}
We address the complementarity of the CERN Large Hadron Collider and an
$e^+ e^-$ Linear Collider in determining SUSY model parameters for a 
particular SO(10) SUSY GUT case study.
\end{abstract}
\maketitle


If weak scale SUSY exists, then it almost certainly should be discovered at the
CERN LHC, where a variety of signatures occur as a result of sparticle
production and subsequent cascade decays\cite{bcpt}. In Ref.~\cite{frank},
several case studies within the mSUGRA model have been performed. 
Although model independent sparticle mass measurements are not typically
possible in a hadron collider environment, many mass 
{\it differences} can be  measured, some with very high precision. 
It was shown that a fit of masses 
and mass differences from an mSUGRA  case study to mSUGRA model parameters 
will allow a determination of 
model parameters to relatively high precision. 

A possible problem is that mSUGRA may not be a valid description of nature.
In this note, we adopt instead a case study based on a SUSY $SO(10)$ GUT with 
a high degree of Yukawa coupling 
unification\cite{bf}. The $SO(10)$
parameters  considered for our case study, as well as a sample of
mass spectra, are shown in TABLE \ref{Table:one}.
The signatures of the model
are very much mSUGRA-like, since $R$ parity is conserved, and the gravitino
decouples. However, $D$-term splitting of scalar masses, $M_D$,  
due to the breakdown of
$SO(10)$ GUT symmetry destroys universality within the generations, and 
leads to
a somewhat different spectrum than is predicted by the mSUGRA model.
In this case, while LHC may discover SUSY, it will remain for a LC to 
provide model independent mass measurements which will reflect the true 
nature of the underlying model.

In our case study, the lightest Higgs scalar $h$ 
should be discovered at the Fermilab Tevatron $p\bar{p}$ collider
if sufficient integrated luminosity is achieved. In addition, clean trileptons
from $\tw_1\tz_2\to 3\ell$ will occur at the 0.65 fb level using cuts
SC2 of Ref. \cite{bdpqt}; this is just above the $3\sigma$ level for
25 fb$^{-1}$ of integrated luminosity, and would give an indication
of the mass difference $m_{\tz_2}-m_{\tz_1}$.

\subsubsection{Mass measurements at the LHC}

The main results for 
the reconstruction of sparticle decay chains, 
and the extraction of certain mass 
differences, were summarized in Ref. \cite{frank}. We rely on these results
when estimating the precision of the determination of mass differences at 
the LHC.

We envision a scenario in which initially certain mass differences and 
later the 
lightest Higgs mass will be measurable at the LHC, providing partial 
information on the SUSY mass spectrum. In particular, based on the $SO(10)$
 model, 
we assume that three mass differences are in the reach of the LHC: 
$\delta_1 = 
m_{\tz_2} - m_{\tz_1}$, $\delta_2 = m_{\tb_1} - m_{\tz_1}$, and 
$\delta_3 = m_{\tg} - m_{\tz_1}$. In the following we 
assess the potential precision of these measurements.

\begin{figure}
\includegraphics[width=6cm,height=6cm]{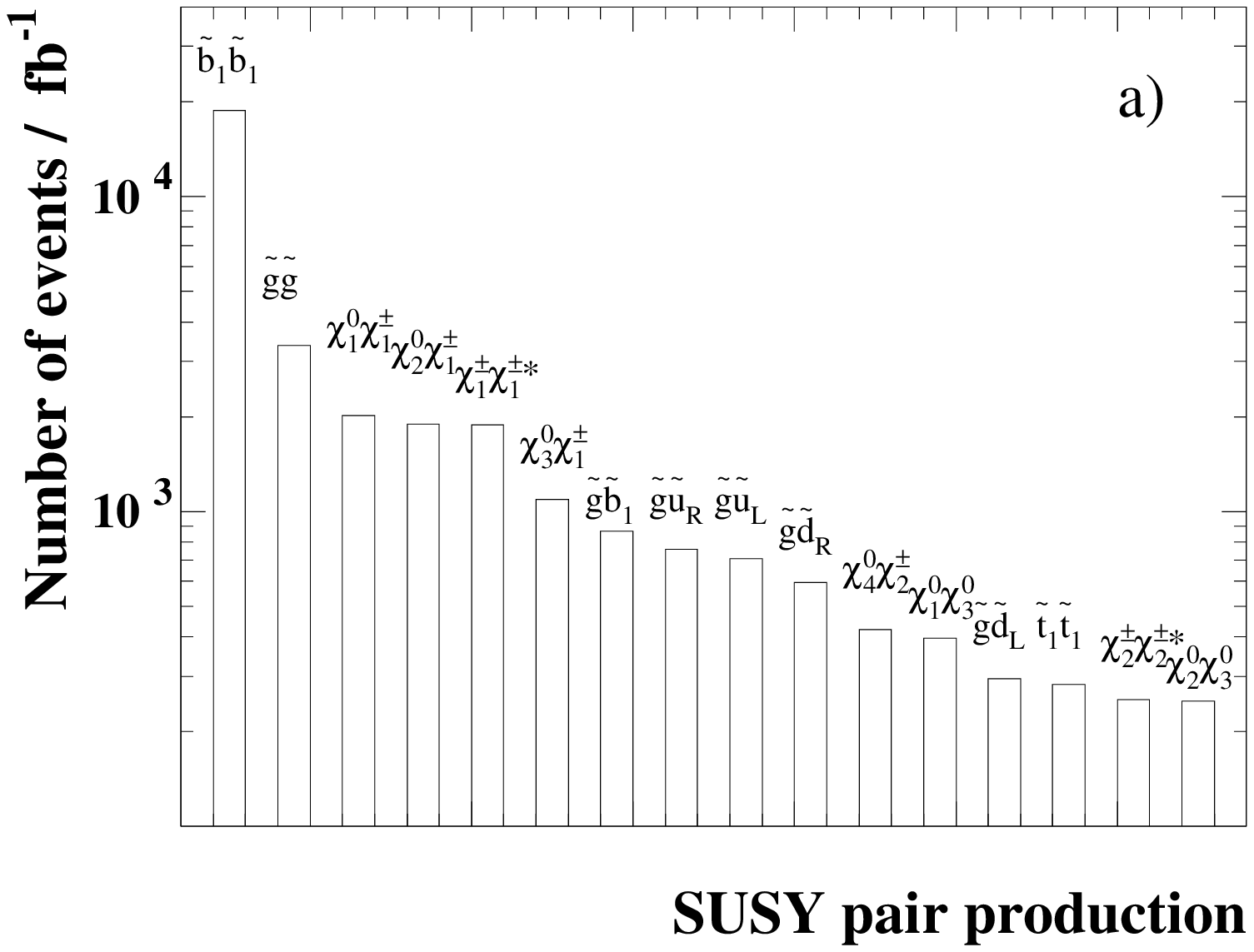}
\includegraphics[width=6cm,height=6cm]{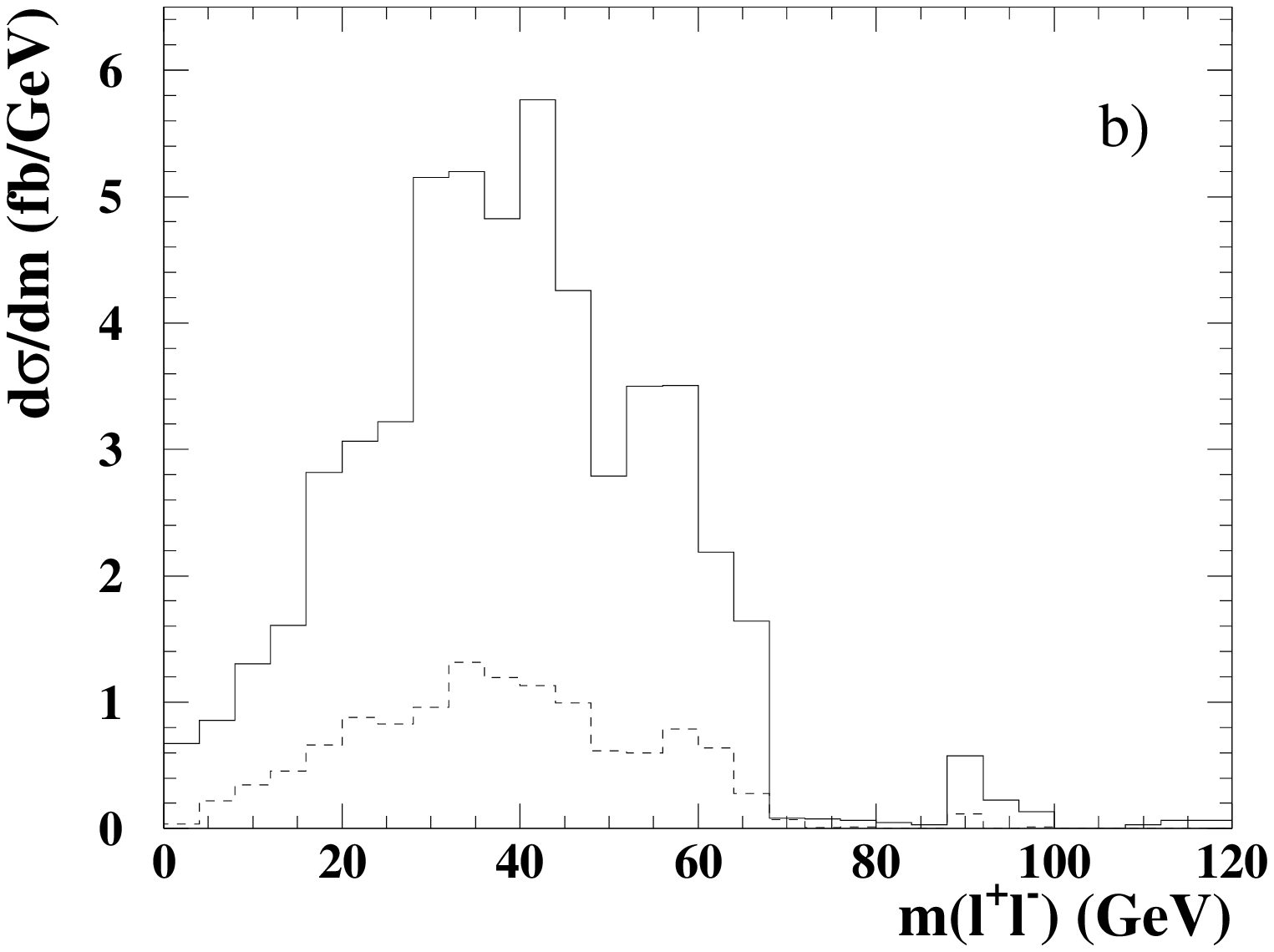}\\
\includegraphics[width=6cm,height=6cm]{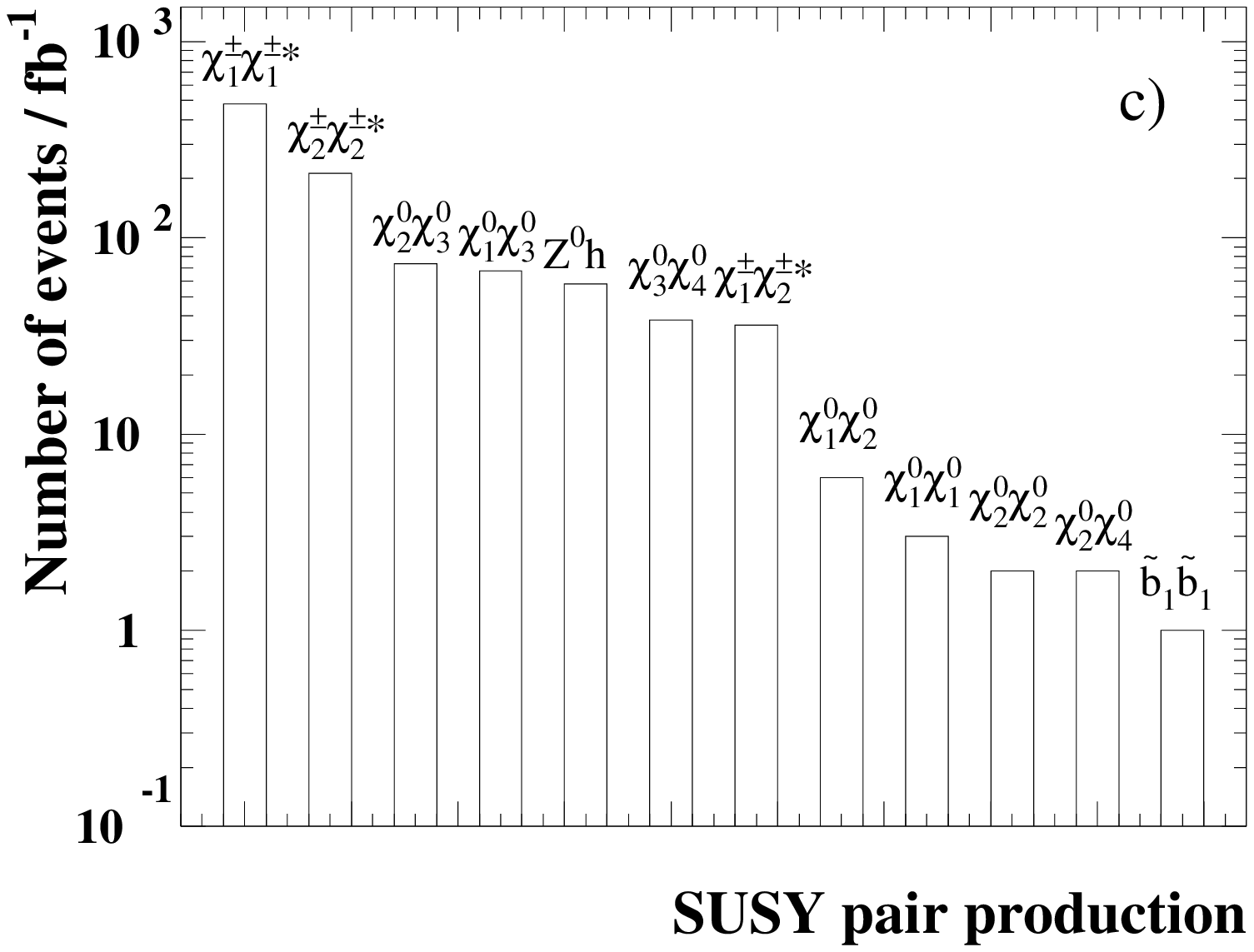}
\includegraphics[width=6cm,height=6cm]{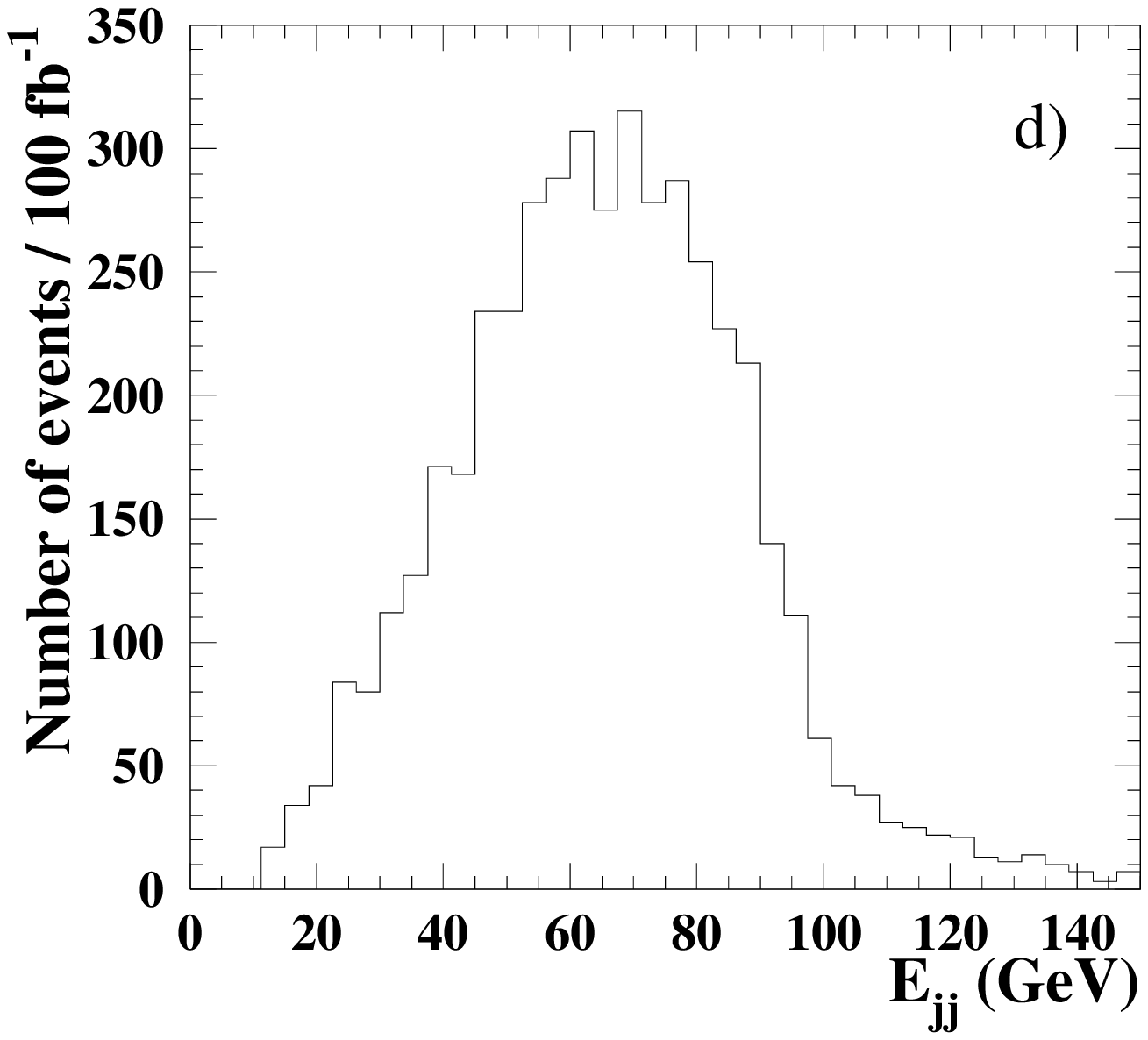}

\caption{}
\label{Fig:one}
\end{figure}

An inspection of the SUSY production cross sections, plotted in 
Fig.~\ref{Fig:one}a, shows that at the LHC the $\tb_1$ pair 
production  rate is the highest, with $\sigma(\tb_1 \tb_1) \sim 1.9 \times 
10^4$ fb, followed by gluino pair production, with $\sigma(\tg \tg) \sim
3.4 \times 10^3$ fb. In our case study, gluinos decay at $\sim 100\%$
to $b\tb_1$. Furthermore, since $\tb_1$ decays
only to  $\tz_i b, i=1,2,3$, and in a proportion 1:1:2 roughly, and 
$\tz_{2,3} \to \tz_1 \ell^+ \ell^-$ branching ratio is a few percent, 
the opposite sign same flavor $\ell^+ \ell^-$ invariant mass 
distribution will have two sharp upper edges.
Indeed in Fig.~\ref{Fig:one}b (although with a rather poor statistics),
one can see two distinct end-points: the first (second) indicates 
$m_{\tz_{2(3)}} - m_{{\tz}_1} =$ 46.1 (71.5) GeV. The solid (dashed)
histogram contains 1049 (961) all SUSY (gluino) pair production events with 
the final state $n$ jets $+ \ell^+ \ell^- + \eslt$, with $n \geq 4$, 
$E_T^{jet} > $  30 GeV and $\eslt >$ 100 GeV.

Although our case study is somehow different from the five points studied
in Ref.~\cite{frank}, we assume here that based on the dilepton mass 
distribution in Fig.~\ref{Fig:one}b similar analysis can be performed and  
$m_{{\tz}_2} - m_{{\tz}_1}$  can be expected to be measured with a 
precision of about 500 MeV. Moreover,  we estimate the precision of the 
measurements $\Delta(\delta_2) = \pm 10$ GeV, and 
$\Delta(\delta_3) = \pm 20$ GeV.  We also assume that after these mass 
differences are extracted, the mass of the lightest Higgs boson will also 
be determined with a precision of 3 GeV.

\subsubsection{Global fit of mSUGRA parameters}

If the LHC determined the above masses to the outlined precision, then it is
possible to perform a global fit to a model to 
determine the basic parameters of the 
underlying theory. Since the signatures are mSUGRA-like, it is natural to first
fit to that particular model.
Using ISAJET 7.58, we begin by generating $1.6\times 10^5$ mSUGRA models, 
uniformly distributed in the parameter space:
\begin{eqnarray}
0 < m_0 < 2000 ~{\rm GeV}, ~~~ 0 < m_{1/2} < 1000 ~{\rm GeV}, 
 ~~~ -2 m_0 < A_0 < 2 m_0, ~~~ 3 < \tan\beta < 50,
\end{eqnarray}
for both signs of $\mu$.
Then, for each of the above models we form
\begin{eqnarray}
\chi^2 = \sum_{i=1}^4 \left( \frac{|\delta_i^{ex}|-|\delta_i^{th}|}
{\Delta(\delta_i)} \right)^2,
\end{eqnarray}
with $\delta_i$, for $i=1,2,3$  being the three measured mass 
differences and $\delta_4 = m_h$. The superscript ``ex'' (``th'') stands 
for the  quantities within $SO(10)$ case study (fitted mSUGRA models). 
Finally, we select the mSUGRA model with the lowest $\chi^2$. To simulate a 
situation in which only partial information is available from the LHC 
experiments, we perform four different fits (Fit 1, ..., Fit 4). 
It is probable  that initially only two of the mass differences will be 
available, so Fit 1 and 2 are only two parameter fits to $\{\delta_1, 
\delta_3\}$ and $\{\delta_1, \delta_2\}$ respectively. Fit 3 fits all 
three mass differences simultaneously, 
and Fit 4 fits the mass differences and the lightest Higgs mass.

The results of the four fits are displayed in TABLE \ref{Table:one}. We 
can see  that even with only two mass differences measured, the $\chi^2$
is low only in Fit 1, which does not depend on the scalar masses. 
This is somehow expected since  the $D$-term present in the  $SO(10)$  
case is responsible for the difference in the scalar  
sector between this model 
and mSUGRA.  With more information accumulating, it will become obvious 
that the model is not mSUGRA, but the vital  point is that it would be 
difficult for the LHC to narrow the masses such that the $SO(10)$ model 
can be singled out.

\subsubsection{The role of the LC}

In our scenario the LHC discovers SUSY, and moreover after years of 
running can give information on mass differences of supersymmetric 
particles. Although  LHC cannot uniquely determine the nature of 
supersymmetry breaking, the combinations of some mass differences 
measured  can  distinguish  our case study  $SO(10)$ model from an
mSUGRA model.

In order to obtain more information on supersymmetric particle masses, 
a precision machine such as a high energy linear collider (LC) will
be required.
In Fig.~\ref{Fig:one}c we show the dominant cross sections within our case 
study at a 500 GeV LC. In particular, the process $e^+ e^- \to \tw_1 \tw_1$ 
leading to 2--jet + $\ell + \eslt$ can  give a very accurate estimates of  
$m_{\tz_1}$ and  $m_{\tw_1}$ through the 2-jet 
energy  distribution \cite{jlc}, 
as shown in Fig.~\ref{Fig:one}d. With the use of  $\tz_1$ mass information 
the LHC data can be re-analyzed  and the correct mass values of the second and 
third lightest neutralinos, gluino and lightest squarks can be obtained. 
In principle all the neutralino and chargino masses can be measured at a
LC,  however to measure the complete SUSY mass spectra the complementary  
role of the  two colliders is necessary.

\begin{table}
\caption{}
\label{Table:one}
\begin{tabular}{lrrrrr}
\hline
               & ~ SO(10) &~~~Fit 1&~~~ Fit 2&~~~ Fit 3&~~~ Fit 4\\
\hline \hline
Parameters fitted  &   -- &
$\delta_{1,3}$ & $\delta_{1,2}$ & $\delta_{1,2,3}$ & $\delta_{1,2,3,4}$ \\
$\chi^2$       &       -- &       1 &     76 &     181 &     228 \\
\hline
$m_{16} (m_0)$ &   1022.0 &  1050.0 &  150.0 &   150.0 &   150.0 \\
$m_{10}$       &   1315.0 &     --  &    --  &     --  &     --  \\
$M_D$          &    329.8 &     --  &    --  &     --  &     --  \\
$m_{1/2}$      &    250.0 &   257.5 &  167.5 &   167.5 &   167.5 \\
$A_0$          &  -1325.0 &   630.0 &  270.0 &   270.0 &   210.0 \\
$\tan\beta$    &     48.0 &    46.5 &   39.4 &    39.4 &    37.1 \\
$\mu$          &   -143.2 &  -148.3 & -208.1 &  -208.1 &  -210.0 \\
\hline
$m_{\tilde g}$ &    649.0 &   644.9 &  420.7 &   420.7 &   421.0 \\
$m_{\tst_1}$   &    530.7 &   768.0 &  306.6 &   306.6 &   303.6 \\
$m_{\tb_1}$    &    239.5 &   829.8 &  302.3 &   302.3 &   310.3 \\
$m_{\tz_1}$    &     85.3 &    85.1 &   62.4 &    62.4 &    62.5 \\
$m_{\tz_2}$    &    131.4 &   130.9 &  109.2 &   109.2 &   109.8 \\
$m_{\tw_1}$    &    119.0 &   118.4 &  108.9 &   108.9 &   109.5 \\
$m_h$          &    119.5 &   113.8 &   91.9 &    91.9 &   106.9 \\
\hline
\end{tabular}
\end{table}

\begin{acknowledgments}
We would like to thank X.~Tata for useful discussions.
\end{acknowledgments}


\end{document}